\documentstyle[preprint,aps]{revtex} 
\input{epsf}
\tighten 
\begin{document}
\preprint{\footnotesize CWRU-P3-96, OUTP-96-07P}
\title{No Crisis for Big Bang Nucleosynthesis}
\author{\large Peter J. Kernan \vspace{5mm}}
\address{Department of Physics, Case Western Reserve University, \\ 
          10900 Euclid Avenue, Cleveland, OH 44106-7079}
\author{\large Subir Sarkar \vspace{5mm}} 
\address{Theoretical Physics, University of Oxford, \\ 
          1 Keble Road, Oxford OX1 3NP \vspace{1cm}}
\date{\sf astro-ph/9603045, revised 28 May 1996} 
\maketitle
\begin{abstract}
Contrary to a recent claim, the inferred primordial abundances of the
light elements are quite consistent with the expectations from
standard big bang nucleosynthesis when attention is restricted to
direct observations rather than results from chemical evolution
models. The number of light neutrino (or equivalent particle) species
($N_\nu$) can be as high as 4.53 if the nucleon-to-photon ratio
($\eta$) is at its lower limit of $1.65 \times 10^{-10}$, as
constrained by the upper bound on the deuterium abundance in high
redshift quasar absorption systems. Alternatively, with $N_\nu = 3$,
$\eta$ can be as high as $8.90 \times 10^{-10}$ if the deuterium
abundance is bounded from below by its interstellar value.
\end{abstract}
\widetext

In a recent Letter, Hata {\it et al.}\ \cite{hata95} have made the
startling claim that the number of light neutrino species deduced from
considerations of big bang nucleosynthesis (BBN) is $N_{\nu} = 2.1 \pm
0.3\ (1\sigma)$, i.e. inconsistent with the standard model
($N_{\nu}=3$) at the $98.6\%$ C.L. Their analysis is based on 3 key
inputs. First, they adopt a primordial $^4{\rm He}$ mass fraction,
$Y_{p} = 0.232 \pm 0.003\ ({\rm stat}) \pm 0.005\ ({\rm syst})$,
estimated \cite{olive95} from selected observations
\cite{pagel92skill93} of low metallicity HII regions in blue compact
galaxies (BCG). Second, they adopt for the primordial abundances of
${\rm D}$ and $^3{\rm He}$, $y_{2p} \equiv {{\rm D}/{\rm H}} =
3.5^{+2.7}_{-1.8} \times 10^{-5}\ (95\%\,{\rm C.L.})$ and $y_{3p} \equiv
{^3{\rm He}/{\rm H}} = 1.2 \pm 0.3 \times 10^{-5}\ (95\%\,{\rm C.L.})$,
using a ``generic'' chemical evolution model \cite{hata96} normalized
to solar system abundances \cite{geiss93}, and convolved with BBN
predictions. Finally, they estimate the primordial abundance of
$^7{\rm Li}$ to be $y_{7p} \equiv {^7{\rm Li}/{\rm H}} =
1.2^{+4.0}_{-0.5} \times 10^{-10} (95\%\,{\rm C.L.})$, and state that
this is consistent with observational data on Pop~II stars, taking
into account possible post big bang production and stellar
depletion. They now compare all four elements simultaneously against
the theoretical predictions, with the theoretical uncertainties
determined by Monte Carlo methods \cite{kernan94}, and obtain the
likelihood function for $N_{\nu}$. This yields the best fit quoted
earlier, corresponding to an upper bound of $N_{\nu} < 2.6\ (95\%\,{\rm
C.L.})$. Hata {\it et al.}\ suggest that this ``crisis'' can be
resolved if the $^4{\rm He}$ mass fraction has been underestimated by
$0.014 \pm 0.004\ (1\sigma)$, or if the constraint on the ${\rm D}$
abundance is relaxed by assuming that the $^3{\rm He}$ survival factor
$g_3$ is $\leq 0.10$ at $95\%\,{\rm C.L.}$ rather than its adopted
value of $0.25$ in their chemical evolution model.

In an accompanying Letter, Copi {\it et al.}\ \cite{copi95a} perform a
Bayesian analysis, adopting the same $^4{\rm He}$ abundance
\cite{olive95} but allowing for a larger statistical error $\sigma_Y$,
and a possible offset in the central value, $Y_p = 0.232 +
\Delta{Y}$. Departing from previous assumptions \cite{copi95b}, they
consider three different models of galactic chemical evolution which
have varying degrees of $^3{\rm He}$ destruction \cite{copi95c} (with
the average survival fraction being about $15\%$ in Model~1, the
extreme case), and two possible values for the primordial $^7{\rm Li}$
abundance estimated from data on Pop~II stars, $^7{\rm Li}/{\rm H} =
1.5 \pm 0.3 \times 10^{-10}$ (no depletion), and $^7{\rm Li}/{\rm H} =
3.0 \pm 0.6 \times 10^{-10}$ (depletion by a factor of 2).They too
conclude that $Y_p$ has been systematically underestimated by about
0.01. Adopting $\Delta{Y} = 0.01$, allowing for maximal $^3{\rm He}$
destruction (Model~1), and assuming zero prior for $N_\nu<3$, then
yields the $95\%\,{\rm C.L.}$ bound $N_\nu < 3.5$ for the lower $^7{\rm
Li}$ abundance; the bound decreases to 3.4 for the higher $^7{\rm Li}$
abundance. Copi {\it et al.}\ \cite{copi95a} suggest that the
situation can be clarified by measuring ${\rm D}$ in quasar absorption
systems (QAS) and testing whether $^7{\rm Li}$ has indeed been
depleted in Pop~II stars.

In fact it has been recognized for some time
\cite{davidson85shields87} that the systematic error in $Y_p$ may be
larger than in the above estimate \cite{olive95}. Recently it has been
emphasized \cite{sasselov94} that previous analyses
\cite{pagel92skill93} have adopted an old set of helium emissivities
\cite{brock72} while the recalculated new set \cite{smits96} allow
much better fits to detailed line ratios. This question has been
examined in a study \cite{izotov9496} of 27 HII regions in 23 BCG
identified in the Byurakan surveys, where, as before, the regression
of the helium abundance against `metals' ($Z$) such as oxygen and
nitrogen is extrapolated down to zero metallicity to extract its
primordial value. Whereas use of the old emissivities yields $Y_{p} =
0.229 \pm 0.004$ in agreement with previous results, use of the new
emissivities (together with new correction factors \cite{kingdon95}
for the collisional enhancement of He~I emission lines) raises the
value to \cite{izotov9496}
\begin{equation}
 Y_{p} = 0.243 \pm 0.003 . 
\label{Ypnew}
\end{equation}
Both the dispersion of the data points in the regression plots and
their slope (${d}Y/{d}Z \approx 1.7 \pm 0.9$) are now smaller than
found before \cite{olive95}, in better agreement with theoretical
expectations. The detailed analysis \cite{izotov9496} suggests a
residual systematic uncertainty in $Y_{p}$ of $\pm0.001$, so we adopt
the $95\%$ C.L. range $0.236 - 0.250$.

Secondly, ${\rm D}$ has already been detected through its
Lyman-$\alpha$ absorption lines in the spectrum of the quasar
Q0014+813, due to a foreground QAS at redshift $z \simeq 3.32$
\cite{cowiecars94}. The derived abundances lie in the range
\begin{equation}
 {\rm D}/{\rm H}~\vert_{QAS (1)} \approx (1.4 - 2.5) \times 10^{-4} .
\label{DLya1}
\end{equation}
Further observations have resolved ${\rm D}$ lines at $z=3.320482$ and
$z=3.320790$, thus eliminating the possibility of confusion with an
`interloper' hydrogen cloud \cite{rughog96a}. The measured abundances
in the two clouds are, respectively, ${\rm D/H} = 10^{-3.73 \pm 0.12}$
and $10^{-3.72 \pm 0.09}$ (where the errors are {\em not} gaussian);
an independent lower limit of ${\rm D/H} \geq 1.3 \times 10^{-4}$ is
set on their sum from the Lyman limit opacity. Recently, there has
been a detection of ${\rm D}/{\rm H} = 1.9^{+0.6}_{-0.9} \times
10^{-4}$ in another QAS at $z=2.797957$ towards the same quasar
\cite{rughog96b}. The errors are higher because the ${\rm D}$ feature
is saturated; nevertheless a $95\%$ C.L. lower limit of ${\rm D}/{\rm
H} > 0.7 \times 10^{-4}$ is set. There have been other, less
definitive, observations of QAS consistent with this abundance,
e.g. ${\rm D/H} \approx 10^{-3.95 \pm 0.54}$ at $z=2.89040$ towards
GC0636+68 \cite{hogan95a}, ${\rm D/H} \lesssim 1.5 \times 10^{-4}$ at
$z=4.672$ towards BR 1202-0725 \cite{wampler95} and ${\rm D/H}
\lesssim 10^{-3.9 \pm 0.4}$ at $z=3.08$ towards Q0420-388
\cite{cars96}. However, very recently, other observers have found much
lower values in QAS at $z=3.572$ towards Q1937-1009 \cite{tytler96}
and at $z=2.504$ towards Q1009+2956 \cite{burles96}; their average
abundance is
\begin{equation}
 {\rm D}/{\rm H}~\vert_{QAS (2)} = 2.4 \pm 0.3\ ({\rm stat}) \pm 0.3\ 
  ({\rm syst}) \times 10^{-5} .
\label{DLya2}
\end{equation}
Unlike the cloud in which the abundance (\ref{DLya1}) was measured,
these QAS also exhibit absorption due to carbon and silicon, whose
synthesis in stars would have been accompanied by destruction of D. It
is argued \cite{tytler96} that this must have been negligible since
the metallicity is very low. Although this is true averaged over the
cloud, large fluctuations in the observed ${\rm D}$ abundance are
possible since the mass of absorbing gas covering the QSO image is
only $\sim10^{-6} M_{\odot}$; thus ${\rm D}$ may well have been
significantly depleted in it by a star which was not massive enough to
eject `metals' \cite{rughog96b}. Keeping in mind that ${\rm D}$ is
{\em always} destroyed by stellar processing, the high ${\rm D}$
measurement (\ref{DLya1}) in the chemically unevolved cloud should be
taken as a conservative upper limit on its primordial abundance. (The
consequences of assuming that this measurement provides the true
primordial value have been investigated elsewhere \cite{krauss94}.)

More data is clearly needed to establish that there is indeed a
``ceiling'' to the ${\rm D}$ abundance. Nevertheless, these
observations already call into question the chemical evolution model
\cite{hata96} employed by Hata {\it et al.} This model assumes that
primordial ${\rm D}$ is burnt in stars to $^3{\rm He}$, a fraction
$g_3 \geq 0.25$ of which survives stellar processing when averaged
over all stars \cite{dearborn86}. However, the recent measurement of
$^3{\rm He}/^4{\rm He} = 2.2 ^{+0.7}_{-0.6} ({\rm stat}) \pm 0.2 ({\rm
syst}) \times 10^{-4}$ in the local interstellar gas \cite{glogei96}
is close to its value of $1.5 \pm 0.3 \times 10^{-4}$ in the pre-solar
nebula \cite{geiss93}, demonstrating that the $^3{\rm He}$ abundance
has not increased significantly in the $4.6 \times 10^9$ yr since the
formation of the solar system. Indeed, $g_3$ must be less than $0.1$
if an initial ${\rm D}$ abundance as high as in Eq.\ (\ref{DLya1}) is
to be reduced to its present value in the interstellar medium (ISM)
\cite{mccull92lin95}
\begin{equation}
 {\rm D}/{\rm H}~\vert_{ISM} \approx 1.5 \pm 0.2 \times 10^{-5} ,
\label{DISM}
\end{equation}
without producing $^3{\rm He}$ in excess of its observed value,
$^3{\rm He}/{\rm H} \approx (1 - 4) \times 10^{-5}$, in galactic HII
regions \cite{balser94}. This would be so if there is net destruction
of ${^3{\rm He}}$ in $(1-2)~M_{\odot}$ stars through the same
mixing process which appears to be needed to explain other
observations, e.g. the $^{12}$C/$^{13}$C ratio \cite{hogan95b}; a
plausible mechanism for this has been suggested recently
\cite{char95}. Note that the ISM abundance of ${\rm D}$ sets a lower
limit to its primordial value $y_{2p}$, since there is no known
astrophysical source of ${\rm D}$ \cite{astroD}.

Finally, accurate ${^7{\rm Li}}$ abundances have been determined for
80 hot, metal-poor Pop~II stars, of which 3 have no detectable lithium
\cite{thor94}. Ignoring these reveals a trend of increasing ${^7{\rm
Li}}$ abundance with both increasing temperature and increasing
metallicity implying that about $35\%$ of the $^7{\rm Li}$ was
produced by galactic cosmic ray spallation processes. The average
value in the hottest, most metal-poor stars is \cite{thor94}
\begin{equation}
 ^7{\rm Li}/{\rm H}~\vert_{Pop~II} = 
  (1.05 - 2.63) \times 10^{-10}\,(95\%\,{\rm C.L.}) . 
\label{Li7II}
\end{equation}
However it may be premature to identify this with the primordial
abundance until the {\em absence} of $^7{\rm Li}$ in some of these
stars is understood. The primordial abundance may instead correspond
to the much higher value observed in Pop~I stars which has been
depleted down to (and in some cases, below) the Pop~II
`plateau'. Stellar evolution modelling \cite{chaboyer94} then
indicates that the initial abundance could have been as high as
\begin{equation}
 ^7{\rm Li}/{\rm H}~\vert_{Pop~I} = 
  (9.54 - 15.1) \times 10^{-10}\,(95\%\,{\rm C.L.}) .
\label{Li7I}
\end{equation}
A recent study \cite{ryan96}, which includes new data on 7 halo
dwarfs, fails to find evidence of significant depletion through
diffusion, although other mechanisms are not excluded. For example,
stellar winds can deplete a primordial abundance of $^7{\rm Li}/{\rm
H} = 10^{-9.5 \pm 0.1}$ down to the Pop II value (\ref{Li7II}) in a
manner consistent with observations \cite{vauchar95}.

\begin{figure}[tbh]
\mbox{\epsfxsize\hsize\epsffile{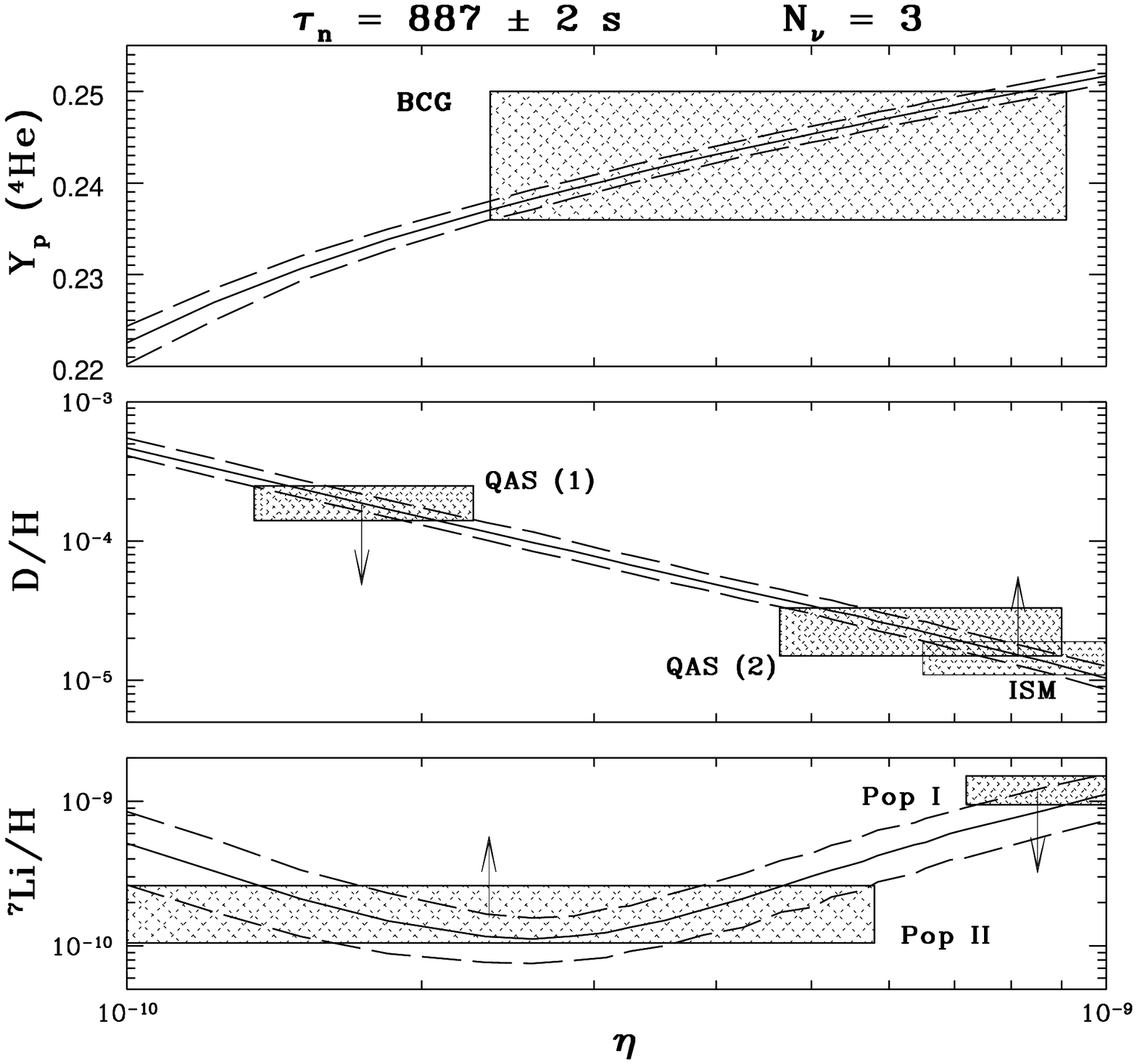}}
\caption{Predicted light element abundances versus $\eta$, with $95\%$
 C.L. limits determined by Monte Carlo. The rectangles indicate the
 various observational determinations. Only the $^4{\rm He}$ abundance
 is established to be primordial in origin.}
\label{abund}
\end{figure}

In Fig.\ \ref{abund} we show that the standard model ($N_\nu = 3$)
{\em can} be consistent with these observations over a wide range of
the nucleon-to-photon ratio, $\eta \equiv n_N/n_\gamma$. It will be
close to its minimum allowed value if $y_{2p}$ is actually given by
Eq.\ (\ref{DLya1}) and $y_{7p}$ by Eq.\ (\ref{Li7II}), while it will
be close to its maximum allowed value if $y_{2p}$ is given by Eq.\
(\ref{DISM}) and $y_{7p}$ by Eq.\ (\ref{Li7I}). Of course a value in
between is also possible, depending on the true primordial abundances
of ${\rm D}$ and $^7{\rm Li}$. Since only the $^4{\rm He}$ abundance
(Eq.\ \ref{Ypnew}) can be reasonably taken to be primordial, we do not
attempt a likelihood analysis. Instead we determine the upper bounds
on the parameters $N_\nu$ and $\eta$ corresponding to the two extreme
possibilities above, taking into account uncertainties in the nuclear
cross-sections and the neutron lifetime by Monte Carlo methods
\cite{kernan94,krauss95}.

\begin{figure}[tbh]
\mbox{\epsfxsize\hsize\epsffile{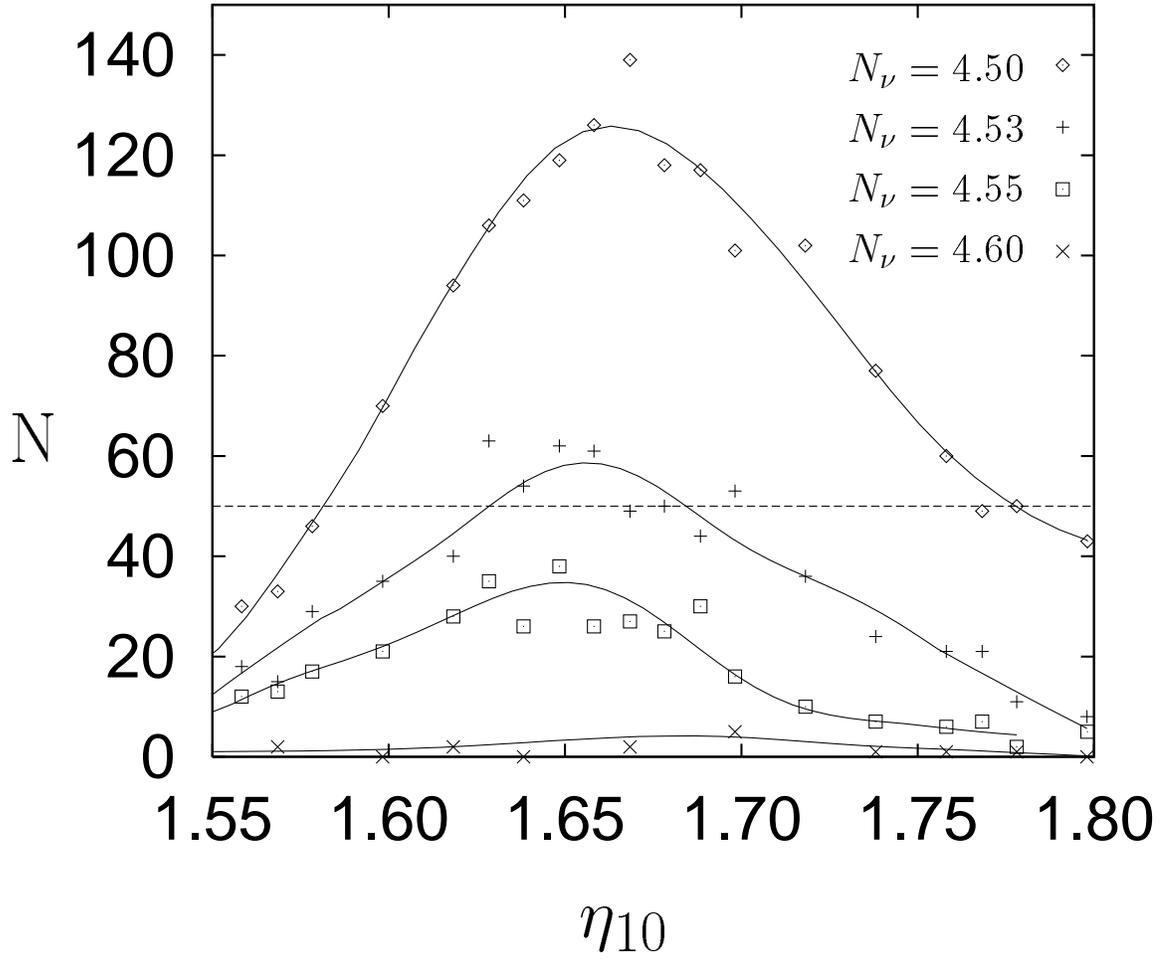}}
\caption{Number of Monte Carlo runs (out of 1000) which simultaneously
 satisfy the constraints $Y_{p} \leq 0.25$, $y_{2p} \leq 2.5 \times
 10^{-4}$ and $y_{7p} \leq 2.6 \times 10^{-10}$, as a function of
 $\eta$ (in units of $10^{-10}$), for various values of $N_{\nu}$.}
\label{Nnumax}
\end{figure}

First we evaluate how many species of light neutrinos are allowed
subject to the upper limits $Y_p < 0.25$ (Eq.\ \ref{Ypnew}), $y_{2p} <
2.5 \times 10^{-4}$ (Eq.\ \ref{DLya1}) and $y_{7p} < 2.6 \times
10^{-10}$ (Eq.\ \ref{Li7II}). Fig.\ \ref{Nnumax} plots the number
($N$) of computation runs (out of 1000) which satisfy the joint
observational constraints, for different values of $N_{\nu}$. It is
seen that (up to $\sqrt{N}$ statistical fluctuations) the ``$95\%\,{\rm
C.L.}$'' bound is $N_{\nu} < 4.53$. This bound varies with the adopted
upper limit to the helium abundance as
\begin{equation}
\label{Nnu4.53}
 N_{\nu}^{max} = 3.75 + 78\ (Y_{p}^{max} - 0.240) .
\end{equation}
For $N_\nu = 3$, $Y_p$ should exceed $0.230$; to have $Y_p$ exceed
0.236 (Eq.\ \ref{Ypnew}), we require $\eta > 2.02 \times 10^{-10}$.

\begin{figure}[tbh]
\mbox{\epsfxsize\hsize\epsffile{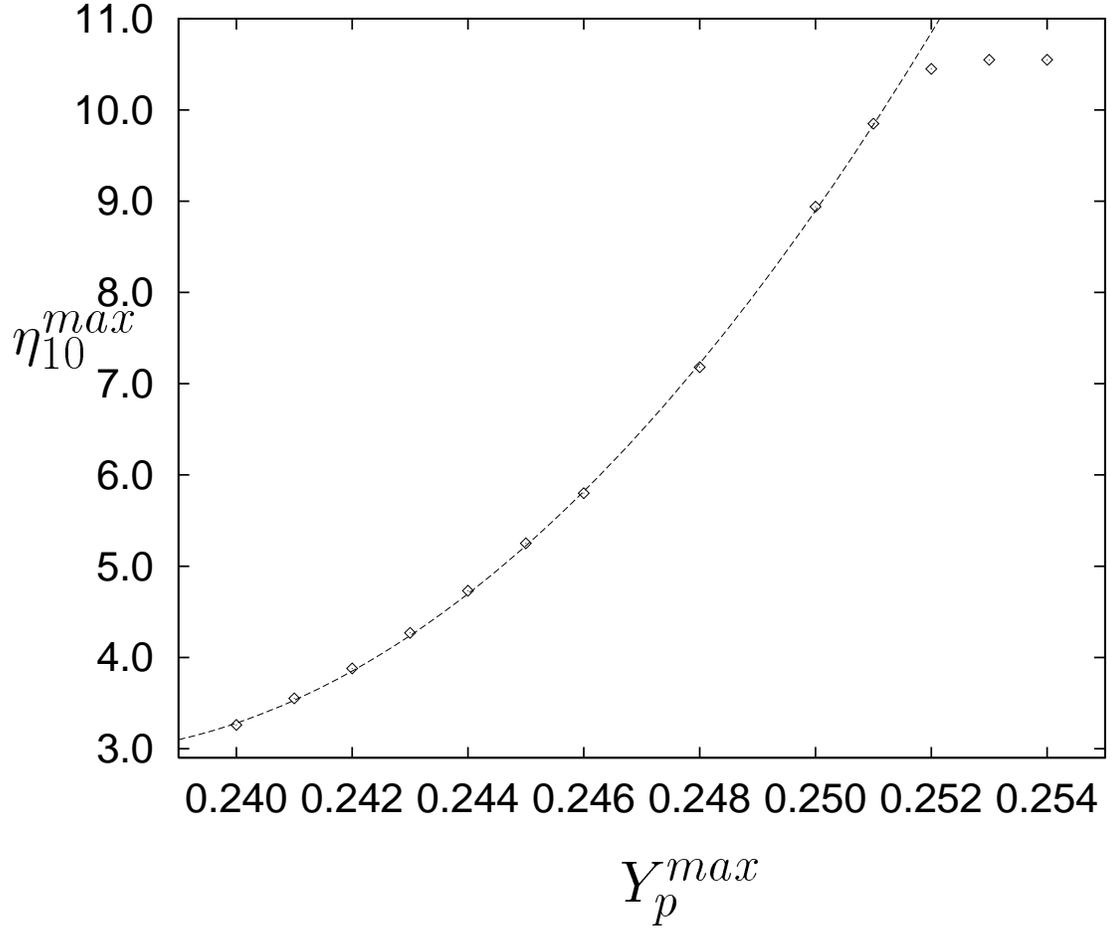}}
\caption{The upper limit to $\eta$ (in units of $10^{-10}$) implied by
 the constraints $y_{2p} > 1.1 \times 10^{-5}$ and $y_{7p} <
 1.5\times10^{-9}$, as a function of the maximum value of $Y_p$.}
\label{etamax}
\end{figure}

Secondly, we calculate the maximum value of $\eta$ permitted by the
data by requiring that 50 runs out of 1000 (up to $\sqrt{N}$
statistical fluctuations) satisfy the constraints
$y_{2p} > 1.1 \times 10^{-5}$ (Eq.\ \ref{DISM}) and
$y_{7p} < 2.6 \times 10^{-10}$ (Eq.\ \ref{Li7II}). A good fit for
$Y_p < 0.247$ is
\begin{equation}
 \eta^{max} = [3.19 + 375.7\,(Y_p^{max} - 0.240)] \times 10^{-10} ; 
\label{etamaxII}
\end{equation}
for higher $Y_p$, the $y_{7p}$ constraint does not permit $\eta$ to
exceed $5.7 \times 10^{-10}$, similar to the result found earlier
\cite{krauss95}. If we choose instead to use the more conservative
limit $y_{7p} < 1.5 \times 10^{-9}$ (Eq.\ \ref{Li7I}), the bound is
further relaxed to
\begin{equation}
 \eta^{max} = [3.28 + 216.4\,(Y_p^{max} - 0.240) + 
               34521\,(Y_p^{max} - 0.240)^2] \times 10^{-10} ,
\label{etamaxI}
\end{equation}
for $Y_p < 0.252$ and saturates at $1.06 \times 10^{-9}$ for higher
values, essentially due to the $y_{2p}$ constraint (see Fig.\
\ref{etamax}). Thus for $Y_p^{max} = 0.25$, we find $\eta^{max} = 8.9
\times 10^{-10}$, which corresponds to a nucleon density in ratio to
the critical density of $\Omega_{N} = 0.033 h^{-2}$, where $h$ is the
Hubble parameter in units of 100 km\,s$^{-1}$\,Mpc$^{-1}$. For
comparison, Hata {\it et al.}\ \cite{hata95} find $\eta =
4.4^{+0.8}_{-0.6} \times 10^{-10} (1\sigma)$ and Copi {\it et al.}\
\cite{copi95a} quote the concordance range $\eta \simeq (2 - 6.5)
\times 10^{-10}$.

In conclusion, the ``crisis'' is not with big bang nucleosynthesis but
with the na\"{\i}ve model of galactic chemical evolution and
restricted set of observational data considered by Hata {\it et al.}\
\cite{hata95}. Their claim that the standard model is excluded follows
from the unreliable ``$95\%\,{\rm C.L.}$'' limits they adopt on the
input elemental abundances. Copi {\it et al.}\ \cite{copi95a} find,
using statistical arguments, that assumptions concerning primordial
abundances in previous work, including their own \cite{copi95b}, must
be relaxed for consistency with the standard model. In fact a variety
of observational data had already pointed to the need for this, as
reviewed elsewhere \cite{sarkar95}.

Our results have important implications for physics beyond the
standard model \cite{sarkar95}. For example, the relaxation of the
bound on the number of neutrino species due to the new ${\rm D}$ and
$^4{\rm He}$ observations now {\em permits} the existence of a gauge
singlet neutrino even if it has large mixing with doublet neutrinos,
as in suggested solutions to the solar and atmospheric neutrino
problems \cite{singlet}. On the other hand, if the primordial ${\rm
D}$ abundance is actually close to its present interstellar value and
the primordial $^7{\rm Li}$ abundance is that inferred from Pop~I
stars, then the nucleon density may be {\em consistent} with the
observed large nucleon fractions in clusters of galaxies, even for a
critical density universe \cite{clustercrisis}.

\acknowledgments{We thank C. Hogan, J. Thorburn and T. Thuan for
communicating their results in advance of
publication. S.S. acknowledges a SERC Advanced Fellowship and support
from the EC Theoretical Astroparticle Network.}

\end{document}